\documentclass[prl,twocolumn,showpacs,preprintnumbers,superscriptaddress,amsmath,amssymb]{revtex4}

\usepackage{epsf,color,graphicx}

\usepackage[dvips]{epsfig}

\begin{document}

\title{Optical bistability in a GaAs based polariton diode}

\author{Daniele Bajoni\footnote{now at CNISM UDR Pavia and Dipartimento di Elettronica, Universit\`{a} degli studi di Pavia, via Ferrata 1, 27100 Pavia, Italy}}
\author{Elizaveta Semenova }
\author{Aristide Lema\^{i}tre}
\author{Sophie Bouchoule}
\author{Esther Wertz}
\author{Pascale Senellart}
\author{Sylvain Barbay}
\author{Robert Kuszelewicz}
\author{Jacqueline Bloch}\email[]{jacqueline.bloch@lpn.cnrs.fr}
\affiliation{CNRS-Laboratoire de Photonique et Nanostructures, Route
de Nozay, 91460 Marcoussis, France}

\date{\today}

\begin{abstract}

We report on a new type of optical nonlinearity in a polariton
\textit{p-i-n} microcavity. Abrupt switching between the strong and
weak coupling regime is induced by controlling the electric field
within the cavity. As a consequence bistable cycles are observed for
low optical powers (2-3 orders of magnitude less than for Kerr
induced bistability). Signatures of switching fronts propagating
through the whole 300 x 300 $\mu m^{2}$ mesa surface are evidenced.

\end{abstract}


\maketitle

After its first observation in a Fabry Perot cavity containing Na
vapor \cite{Gibbs}, optical bistability has been widely explored
in solid state systems for its possible application in all optical
circuits and optical computing \cite{Gibbsbook}. A common approach
is the use of a microcavity in which the resonance frequency
depends on the stored optical energy: optical $\chi^{(3)}$
nonlinearities, of electronic or thermal origin, have been used to
obtain bistability in 1-dimensional \cite{Oudar,Debray} and 2
dimensional \cite{Notomi,Yacomotti} photonic devices, with
switching incident powers around 1 kW/cm$^2$. When part of a
spatially extended bistable system is switched from one stable
state to the other, a front is formed between spatial regions in
different states. If this front is locked, spatial solitons can be
observed \cite{solitons,solitonlaser}, otherwise the front
propagates along the surface until the whole sample has switched
state
\cite{Kuszelewicz}.\\
Recently optical bistability of microcavity polaritons has been
theoretically proposed to generate propagation of switching fronts
which can be used for all optical computation \cite{Kavokin}.
Polaritons are mixed exciton-photon quasi-particles resulting from
the strong coupling regime of excitons with a resonant cavity mode
\cite{Weisbuch}. Polariton-polariton scattering gives rise to giant
$\chi^{(3)}$-type nonlinearities \cite{SavvidisStevenson, Tignon}
which have been recently shown to generate optical bistability
\cite{Bramati}; indications of spatial solitons were also reported
\cite{Weiss}. Another approach for optical bistability is to use the
switch from strong to weak coupling regime due to exciton bleaching
at high pumping power \cite{Houdre,Butte}. This method has been
theoretically proposed in 1996 \cite{Tredicucci}, and some
experimental indication has been reported in 2004
\cite{Gibbspolariton}.\\
\begin{figure}[b]
\includegraphics[width= 0.8\columnwidth]{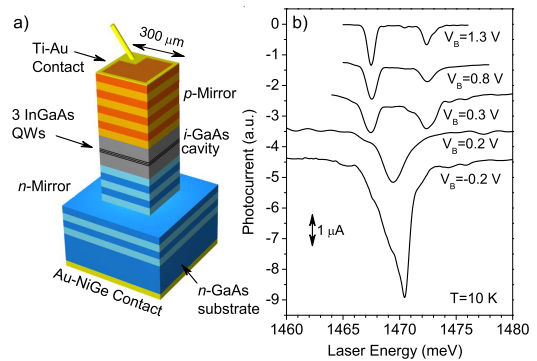}
\caption{(Color online) a) Schematic sample structure.
b)Photocurrent as a function of the laser energy measured with an
optical power P = 10 W/cm$^2$ and for different voltage bias.
Spectra have been vertically shifted for clarity}\label{Fig1}
\end{figure}
In this work, we experimentally demonstrate low-power optical
bistability based on a new non-linear mechanism. Switching between
strong and weak coupling regime is induced controlling the internal
electric field of a \textit{p-i-n} microcavity. Well defined
hysteresis cycles are observed both scanning the external bias or
the optical power.  A model including the changes of optical and
electronic properties between the strong and weak coupling regime is
developed and gives a good overall description of the observed
cycles. Finally we show that a local excitation can produce
commutation of the whole
mesa.\\
The sample (see Fig. \ref{Fig1} a)) is described in details in ref.
\cite{polaritonLED}. Grown on an \textit{n-}doped GaAs substrate, an
undoped GaAs cavity containing 3 In$_{0.05}$Ga$_{0.95}$As QWs is
surrounded by a \textit{p-} doped and an \textit{n-}doped
Ga$_{0.9}$Al$_{0.1}$As/Ga$_{0.1}$Al$_{0.9}$As Bragg mirror. 300 $\mu
m$ square mesas, etched down to the substrate, were connected with
metal contacts. The diode under study, maintained at 10 K, presents
a zero detuning between the cavity mode at normal incidence and the
QW exciton
resonance.\\
\begin{figure}[b]
\includegraphics[width= 0.4\textwidth]{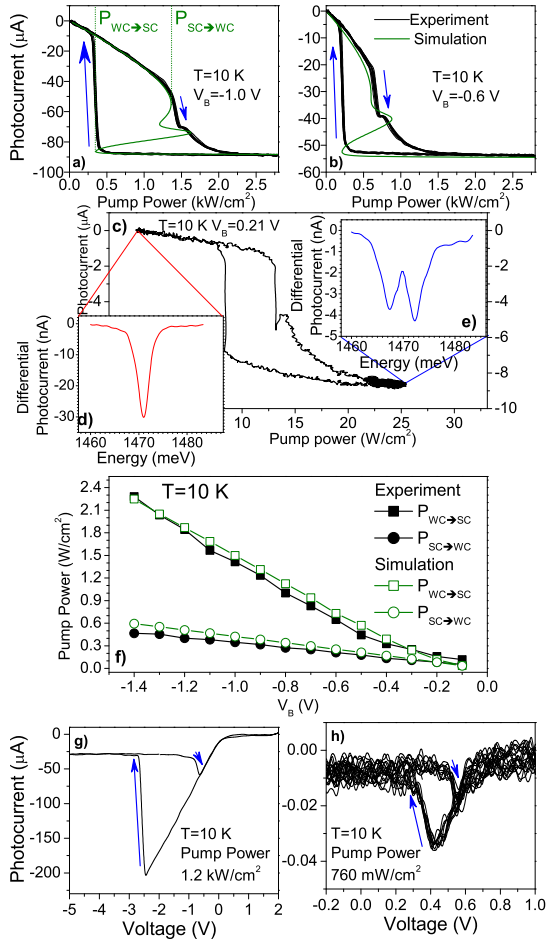}
\caption{(Color online) a) and b) experimental (black) and simulated
(green) photocurrent as a function of the excitation power for to
values of the bias. c) Photocurrent as a function of excitation
power for $V_B=$ 0.22 V. d) and e): Differential photocurrent
spectra measured for $V_B=$ 0.22 V with a chopped probe
($P_{probe}\sim$5 W/cm$^2$), d) $P=0$ and e) $P=$ 25 W/cm$^2$. f)
Experimental (full symbols) and simulated (open symbols) pump powers
for the SC $\rightarrow$ WC and WC $\rightarrow$ SC. g) and h)
photocurrent measured varying the bias at fixed excitation
power.}\label{Fig2}
\end{figure}
Let us first describe the two operating regimes of our bistable
structure. Fig. \ref{Fig1} b) summarizes photocurrent spectra under
normal incidence measured for different values of the external bias
$V_B$. For $V_B > 0.3$ V, two distinct photocurrent peaks are
resolved, attributed to the upper and lower polariton branches
(polariton dispersion has been fully characterized in ref.
\cite{polaritonLED}): the system operates in the strong coupling
(SC) regime \cite{Weisbuch}. On the opposite, below 0.2 V
photocurrent spectra exhibit only one peak (the cavity mode): the
sample has abruptly switched to the weak coupling (WC) regime. To
understand this behavior, the electric field  E at the QW position
has to be considered:
\begin{equation}
E=\frac{\phi_i-V_B+ZI}{L_{cav}} \label{eq1}
\end{equation}
where $\phi_i\simeq$1.48 V is the built in potential, $I$ the
current, $Z$ the load resistance, and $L_{cav} = 237.6$ nm the
intrinsic region thickness. In the measures of Fig. \ref{Fig1} b),
$Z= 50$ $\Omega$ is the coupling impedance of the voltmeter and $ZI$ can be neglected.\\
For $V_B$=0.2 V, $E$ lies around 55 kV/cm so that the Stark effect
is too small to induce the loss of SC regime \cite{SkolnickStark}.
The abrupt switch between SC and WC regime can be understood
considering the enhanced tunneling of electrons and holes out of the
QW in presence of the electric field. The tunneling time
$\tau_{tunnel}^i$ ($i=e,h$) is given by \cite{Bastard}:
\begin{equation}
\tau_{tunnel}^i=\frac{2m^*_iL_{QW}^2}{\hbar\pi}\exp\left(\frac{4}{3\hbar
e E}\sqrt{2m^*_iU^3_i}\right) \label{eq2}
\end{equation}
where $m^*_i$ and $U_i$ are respectively the effective mass and
confinement barrier of electrons and holes, $L_{QW}=8$ nm the QW
width and $e$ the electron charge. In the present QWs the barrier
height is rather small : $U_e=$42 meV, $U_h=$5 meV. For $E=55$
kV/cm equation \ref{eq2} yields $\tau_{tunnel}^h\sim$0.15 ps and
$\tau_{tunnel}^e\sim$0.4 ps, so that $\tau_{tunnel}^h$ is
comparable to the Rabi oscillation period ($\sim$0.12 ps). The
sample switches from SC to WC because the exciton is destroyed by
hole tunneling out of the QW before a Rabi oscillation can be
completed. Notice that the physics governing the switch is
completely different than in previous reports
\cite{Gibbspolariton}. Instead of bleaching the exciton by coulomb
interaction screening and phase space filling \cite{Houdre,Butte},
here the light matter interaction is modified by the internal
electric field. This mechanism is more similar to the
self-electro-optic effect device (SEED) \cite{Chemla1,Couturier},
in which the electric field changes the QW absorption by
quantum-confined Stark effect.\\
The abrupt change in reflectivity described above can be exploited
to generate optical bistability. Let us consider experiments where
the laser is resonant to the lower polariton energy and its power
$P$ is modulated by a symmetric triangular wave. It is focused into
a 50 $\mu$m diameter spot with a $10^{\circ}$ angular aperture. The
photocurrent is measured with an oscilloscope across a $Z=$10
k$\Omega$ load resistance. In Fig. \ref{Fig2} a), b) and c), $V_B$
is constant and chosen so that in the dark the sample operates in
the WC regime. Clear hysteresis cycles are observed in the
photocurrent. Indeed, when $P$ is increased starting from zero
(forward ramp), the sample is initially in the WC regime and the
reflectivity dip lies at the energy of the uncoupled cavity mode so
that the laser is out of resonance. The photocurrent (proportional
to $P$), being negative, progressively reduces $E$ through the $ZI$
term of eq. \ref{eq1}. Eventually $E$ becomes low enough for the
sample to switch into the SC regime. As a result, the laser becomes
resonant and an increased power is transmitted with a further
reduction of $E$, providing a positive feedback loop. Notice that
once in the SC regime, the current tends to saturate; when $ZI$
approaches the value $V_B-\phi_i$ (see eq \ref{eq1}), $E$ tends to
zero and the tunneling times out of the QW (given by eq. \ref{eq2})
become much greater than the radiative recombination time. In this
regime increasing the pumping power increases the flux of emitted
photons rather than the photocurrent. When $P$ is reduced, starting
from the SC regime (backward ramp), the initial transmission of the
laser is high. Thus, for the same value of $P$, the photocurrent is
more intense and  $E$ is lower than in the forward ramp. This means
that the switch occurs at a lower $P$, thus opening the bistable cycle.\\
To support this interpretation, pump and probe experiments were
performed. The sample was biased at $V_B=0.21$ V, and two collinear
beams were used: a constant pump and a chopped tunable probe (with
$P_{probe}=$ 5 W/cm$^2$). The photocurrent was then filtered with a
lock-in amplifier to isolate the differential signal generated by
the probe.  Such differential photocurrent spectra on each side of
the hysteresis cycle are shown in Fig. \ref{Fig2} d) and Fig.
\ref{Fig2} e) (corresponding to P = 0 and 25 W/cm$^2$). The
transition from a single photocurrent peak (WC
regime) to a double peaked spectrum (SC regime) is clearly evidenced.\\
The observed bistability cycles at fixed bias have been reproduced
using a phenomenological model. We describe the transition from SC
to WC driven by the electric field considering the associated
changes in several physical quantities and in carrier dynamics.
Across the transition, the cavity transmission $T$, the conversion
efficiency of photons into carriers
 $\eta$, and the radiative lifetime $\tau_{rad}$ vary
between their WC and SC values. To simulate the transition from WC to SC, transmission
 spectra were computed for increasing values of the exciton dephasing time
 ($\tau_X$). The transfer matrix method \cite{Tredicucci,transferM} was used, solving Maxwell equations
  in each layer
  and imposing continuity conditions at each interface. The evolution with $\tau_X$ of the calculated
  transmission coefficient at the energy of the lower polariton is well reproduced using:

\begin{equation}
T=T_{SC}+\frac{T_{WC}-T_{SC}}{1+\beta
*\tau_X}\label{eqR}
\end{equation}

with $\beta=4.7\times 10^{11}$ s$^{-1}$. $T_{WC}$=0.01, while in
the SC regime the cavity transmission under normal incidence
amounts to 65\%. However due to the $10^{\circ}$ angular aperture
of the excitation beam, the polariton resonance acts as an angular
filter: we estimate that $T_{SC}=0.02$. To model the transition
for $\eta$ and $\tau_{rad}$, the same dependence on $\tau_X$ as in
eq. \ref{eqR} is used. In WC regime $\eta_{WC}$= 0.02 corresponds
to the 2\% QW absorption. In SC regime all photons entering the
cavity become polaritons and $\eta_{SC}$= 1. In WC regime
$\tau_{rad}^{WC}$=200 ns is the carrier radiative recombination
time for a density around 10$^9$ cm$^{-2}$ \cite{Matsusue}. In SC
regime
$\tau_{rad}^{SC}$=2 ps is the polariton recombination time, given by twice the cavity lifetime. \\
For a given photocurrent, the electric field $E$ is given by eq.
\ref{eq1}. The exciton dephasing time is therefore driven by the
competition between phonon dephasing and tunneling out of the QW:
\begin{equation}
\tau_X(E)=\left[\frac{1}{\tau_{X0}}+\frac{1}{\tau_{tunnel}^e(E)}+\frac{1}{\tau_{tunnel}^h(E)}\right]^{-1}\label{eqtauX}
\end{equation}
where $\tau_{X0}= 10$ ps is the exciton dephasing time due to
interaction with phonons for $E=0$. Under steady state conditions
the photocurrent $I$ is linked to the population of electrons and
holes through:
\begin{equation}
I=I_e+I_h=e\left[\frac{n_e(E)}{\tau_{tunnel}^e(E)}+\frac{n_h(E)}{\tau_{tunnel}^h(E)}\right]
\label{eqI}
\end{equation}
where $e$ is the elementary charge. Moreover since electrons and
holes injected by the laser can either recombine radiatively with
decay time $\tau_{rad}(E)$ or tunnel out of the QW, we get:
\begin{equation}
n_i(E)=\eta(E)T(E)\frac{P}{\hbar\omega}\frac{\tau_{rad}(E)\tau^i_{tunnel}(E)}{\tau_{rad}(E)+\tau^i_{tunnel}(E)}
\label{eqpop}
\end{equation}
where $\hbar\omega$=1467.5 meV is the laser energy.\\
From equations \ref{eqI} and \ref{eqpop}, we deduce the explicit
dependence of the pump intensity on $I$ :
\begin{equation}
P=\frac{\hbar\omega I}{e\eta T}
\frac{\left(\tau_{rad}+\tau^e_{tunnel}\right)\left(\tau_{rad}+\tau^h_{tunnel}\right)}{\tau_{rad}\left(2\tau_{rad}+\tau^e_{tunnel}+\tau^h_{tunnel}\right)}
\label{eqP}
\end{equation}
A good overall agreement is found between simulated and
experimental curves as shown in Fig. \ref{Fig2} a) and b)
\cite{footnote}. The width of the hysteresis cycles is well
reproduced. Notice that there is a small discrepancy in the
additional feature that appears in the experiment as a small bump
at $I=-70$ $\mu$A in Fig. \ref{Fig2} a) or at $I=-40$ $\mu$A in
Fig. \ref{Fig2} b). The simulated curves also present a feature at
these current values, coming from the interplay between the
increase of $T$ and of $\tau^i_{tunnel}$ during the switch.
However a detailed microscopic model is needed to obtain a
complete agreement. We define $P_{WC\rightarrow SC}$ and
$P_{SC\rightarrow WC}$ as the pump powers of the switch between WC
and SC and vice versa. Simulated and experimental values of these
two quantities are summarized in Fig.\ref{Fig2} f). The very good
agreement between the model and the experiment further support the
interpretation of the observed cycles
in terms of switching between SC and WC regime.\\
The sample can also be driven varying the bias at fixed $P$: similar
bistable cycles are observed as shown in Fig. \ref{Fig2} g) and h).
In these conditions, bistability  has been observed for P as low as
0.7 W/cm$^2$ (Fig. \ref{Fig2} h)). Although still greater than the
pumping powers used
 in SEEDs (in which bistable cycles have been observed for optical powers as low as 100 $\mu$W/cm$^2$ \cite{Couturier}),
 our device is more
  than two
order of magnitude more efficient than previously reported cavity
based devices \cite{Oudar,Debray,Notomi,Yacomotti}. It is indeed the
first demonstration of low power optical bistability for
polaritonics \cite{Bramati,Gibbspolariton,Kavokin}.

Bistability curves have been observed tuning the laser to both the
lower and upper polariton branches. On the contrary, when tuning the
laser to the cavity mode energy, the feedback between the
photocurrent and the laser transmission is negative, and no
bistability was observed. The sample switching time between the two
states could not be measured : the present diodes have a cut-off
frequency of $\sim$1 kHz due to their large areas. This effect
should however be greatly reduced by patterning the diodes in the
form of micron-sized micropillars \cite{Blochpillars}.\\
\begin{figure}[b]
\includegraphics[width= 0.8\columnwidth]{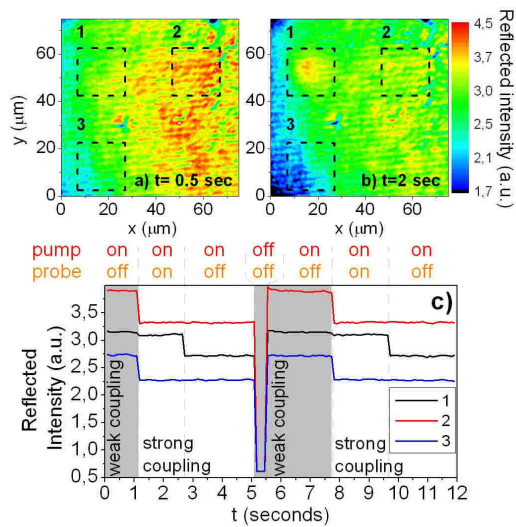}
\caption{(Color online)Real space image (in logarithmic color scale)
of the pump spot reflected on the sample surface. $V_B= -2 V$ and
$P_{pump} \sim$ 2 kW/cm$^2$ setting the system in the bistable
region. a) the sample is in WC regime; b) similar than a) but with
the writing laser beam shone in region 1: the whole surface has
switched into the WC regime; c)reflected intensities averaged on
region 1, 2 and 3 during the sequence indicated on top.}\label{Fig3}
\end{figure}
Let us finally consider regions prepared in different states and
address the existence and the propagation of switching fronts
connecting them. The sample was homogeneously illuminated by a laser
tuned to the lower polariton energy. A spatial image of the
reflected light intensity is shown in Fig. \ref{Fig3} a). P $\sim$ 2
kW/cm$^2$ and $V_B=-2 V$ were chosen to set the sample in the
bistable region, initially in the WC state. A writing laser beam was
focused on a $\sim$ 20 $\mu$m spot in region 1 within the pump spot.
Its intensity is high enough to switch the sample to the SC state.
As shown in Fig. \ref{Fig3} b), the writing beam induces a
reflectivity drop on the whole surface. This is further illustrated
in Fig. \ref{Fig3} c) where the reflected intensity averaged over
each of the three regions labelled 1,2 and 3 are summarized. Regions
2 and 3 are 40 $\mu$m apart from region 1. Initially, the whole
sample is in the WC and presents a high reflectivity. At $t=1.1$ s
the writing beam is turned on and all three regions switch to the SC
state with lower reflectivity (this is not visible in region 1
because the reflectivity drop is compensated by the reflected
writing beam). At $t=2.7$ s the writing beam is turned off and all
regions remain in the same state. The pump is turned off at $t=5.1$
s and on again at $t=5.5$ s: the sample has come back to the WC
state and the previous sequence is repeated. These measurements
highlight that commuting a small area within the pump spot induces a
switch of the whole excited region. This necessarily occurs through
the propagation of a switching front across the sample
\cite{Kuszelewicz}. In our case, it propagates over distances of
several tens of
microns.\\
In conclusion, we have demonstrated pronounced optical bistability
in a \textit{p-i-n} microcavity based on a new electro-optic
mechanism. Driving the internal electric field within the cavity
produces a strong non-linearity with a positive feed back, due to
abrupt switching between strong and weak coupling regime. Signatures
of switching fronts propagating over long distances are evidenced.
Such fronts could in the future be guided by lateral patterning of
the diodes into 1D structures. This opens the way toward the
realization of new polariton based optical logic gates as proposed
in ref. \cite{Kavokin}. Devices with fast switching times could be
achieved by lateral processing of the microcavity into micron sized
pillars. The observed low-power bistability relies on the bleaching
of the strong coupling regime due to the internal electric field,
and could easily be transferred into large band gap
materials \cite{Semond} for room temperature operation.\\
This work was funded by "C'nano Ile de France" and "Conseil
G\'en\'eral de l'Essonne". We thank Paul Voisin and Christophe Minot
for fruitful discussions.


\begin{thebibliography}{99}

\bibitem{Gibbs}
H. M. Gibbs, S. L. McCall, and T. N. C. Venkatesan, Phys. Rev.
Lett. \textbf{36} 1135 (1976).

\bibitem{Gibbsbook}
H. M. Gibbs, \textit{Optical Bistability: Controlling Light with
Ligth} (Academic, New York, 1985).

\bibitem{Oudar} H. M. Gibbs \textit{et al.}, Appl. Phys. Lett. \textbf{41}, 221
(1982). O. Sahlen \textit{et al.}, Ibid. \textbf{50}, 1559 (1987).

\bibitem{Debray}
R. Kuszelewicz \textit{et al.}, Appl. Phys. Lett. 5\textbf{3},
2138 (1988).

\bibitem{Notomi}
T. Tanabe \textit{et al.}, Opt. Lett. \textbf{30} 2575
(2005).

\bibitem{Yacomotti}
A. M. Yacomotti \textit{et al.}, Appl. Phys. Lett. \textbf{88}, 231107 (2006).

\bibitem{solitons}
M. Tlidi, P. Mandel, and R. Lefever, Phys. Rev. Lett. \textbf{73},
640 (1994).

\bibitem{solitonlaser}
Y. Tanguy et al., Phys. Rev. Lett. 100, 013907 (2008)

\bibitem{Kuszelewicz}
I. Ganne et al., Phys. Rev B \textbf{63}, 075318 (2001).

\bibitem{Kavokin}
T. C. H. Liew, A. V. Kavokin, and I. A. Shelykh, Phys. Rev. Lett
\textbf{101}, 016402 (2008).

\bibitem{Weisbuch}
C. Weisbuch et al., Phys. Rev. Lett. \textbf{69} 3314 (1992).

\bibitem{SavvidisStevenson}
P.~G. Savvidis \textit{et al.}, Phys. Rev. Lett. \textbf{84}, 1547
(2000); R.~M. Stevenson \textit{et al.}, Ibid. \textbf{85}, 3680
(2000).

\bibitem{Tignon}
C. Diederichs \textit{et al.}, Nature \textbf{440}, 904 (2006).

\bibitem{Bramati}
A. Baas \textit{et al.}, Phys. Rev B \textbf{70}, 161307(R)
(2004).

\bibitem{Weiss}
Ye. Larionova, W. Stolz, and C. O. Weiss, Opt. Lett. \textbf{33}
321 (2008).

\bibitem{Houdre}
R. Houdr\'{e} \textit{et al.}, Phys. Rev. B \textbf{52} 7810
(1995).

\bibitem{Butte}
R. Butt\'{e} \textit{et al.}, Phys. Rev. B \textbf{65}, 205310 (2002).

\bibitem{Tredicucci}
A. Tredicucci \textit{et al.}, Phys. Rev. A \textbf{54}, 3493
(1996).

\bibitem{Gibbspolariton}
M Gurioli et al., Semicond. Sci. Technol. \textbf{19}, S345 (2004).

\bibitem{polaritonLED}
D. Bajoni \textit{et al.}, Phys. Rev. B \textbf{77}, 113303 (2008)

\bibitem{SkolnickStark}
T. A. Fisher \textit{et al.},
Phys. Rev. B \textbf{51}, 2600 (1995)

\bibitem{Bastard}
G. Bastard, J. A. Brum, and R. Ferreira, Solid State Phys. vol.
\textbf{44}, p. 229, Academic press (1991).

\bibitem{Chemla1}
D. A. B. Miller \textit{et al.}, Appl. Phys. Lett. \textbf{45}, 13
(1984).

\bibitem{Couturier}
J. Couturier, J.C. Harmand, and P. Voisin, Semicond. Sci. Technol.
\textbf{10}, 881 (1995)

\bibitem{transferM}
See, for example, A. Yariv, and P. Teh, \emph{Optical Waves in
Crystals} (Wiley, New York, 1984).

\bibitem{Matsusue}
T. Matsusue and H. Sakaki, Appl. Phys. Lett. \textbf{50}, 1429
(1987).

\bibitem{footnote}
To fit the data, $\phi_i=0.07$ V had to be phenomenologically
imposed in the model. A complete microscopical description is
needed to explain this discrepancy.


\bibitem{Blochpillars}
J. Bloch \textit{et al.}, Superlatt. Microstruct. \textbf{22}, 371
(1997).

\bibitem{Semond}
F. Semond \textit{et al.},  Appl.
Phys. Lett. \textbf{87}, 021102 (2005).

\end{thebibliography}
\end{document}